\documentclass[a4paper,10pt]{article}
\usepackage[utf8x]{inputenc}
\usepackage{amssymb}
\title{ Perturbative Quantum Gravity on Complex Spacetime}
\author{Mir Faizal\\ Department of Mathematics,  Durham University,\\
 Durham, DH1 3LE,  United Kingdom,\\ faizal.mir@durham.ac.uk}

\begin{document}

\maketitle

\begin{abstract}
In this paper we will study non-anticommutative perturbative quantum gravity on 
 spacetime with a complex metric. 
After  analysing the BRST symmetry of this  non-anticommutative perturbative quantum gravity,
 we will also analyse the effect of 
shifting all the fields. We will construct a Lagrangian density which is invariant 
under  the original BRST transformations and the shift transformations in the  Batalin-Vilkovisky   (BV) formalism. 
Finally, we will show that the sum of  
the gauge-fixing term and the ghost term for this shift symmetry invariant Lagrangian density  
 can be elegantly written  down in  superspace with a single Grassmann parameter. 
\end{abstract}

\section{Introduction}

Noncommutative field theory arises in string theory due to  presence of the $NS$
 antisymmetric tensor background \cite{2pas}-\cite{3p}.
However, other backgrounds like the $RR$ background can generate non-anticommutative field theory \cite{ov,se}.
Quantum gravity on noncommutative spacetime has been thoroughly studied \cite{21p}-\cite{23paa}. 
In fact, it is hoped that noncommutativity predicts the existence of the cosmological 
constant which is of the same order as the square of the Hubble's constant
\cite{24p}.  
Perturbative quantum gravity on noncommutative spacetime has been already analysed  \cite{29p}. 
It was found that the graviton propagator was  the same as that in the commutative case. However, 
the noncommutative nature of spacetime was experienced at the level of interactions. 

The idea of noncommutativity of spacetime has been generalized to non-anticommutativity.
In addition to this, quantum field theory has  been studied on  
 non-anticommutative spacetime   
 \cite{k0}. Non-anticommutativity of spacetime occurs if the metric is complex.
 Spacetime with complex metric has
been studied as an interesting example of nonsymmetric gravity \cite{k1}-\cite{ k3}.
Even though nonsymmetric quantum gravity was initially studied in an attempt to unify 
electromagnetism and gravity \cite{k4, k5}, it is now mainly studied  due to its relevance to 
string theory \cite{k6}-\cite{k8}.  Quantum gravity on this 
 non-anticommutative complex spacetime  has also been discussed  before \cite{299p}.

In this paper we will discuss the perturbative quantum gravity on this non-anticommutative 
 complex  spacetime. 
We will first analyse the BRST symmetry of this theory and then study its the
 invariance  under the original
BRST and shift transformations in the BV-formalism. 
Consequently, we will express our results in superspace formalism.

The BRST symmetry 
for  Yang-Mills theories  \cite{30p}-\cite{33pa} 
and spontaneously broken gauge theories \cite{33pb}
 has already been analysed in noncommutative spacetime.
The invariance of a theory, under the original BRST transformations 
and shift transformations, that occurs naturally in 
background field method  can be analysed in the 
BV-formalism  \cite{1}-\cite{3f}. 
Both  the BRST formalism  \cite{4,5} and   the BV-formalism can be given 
geometric meaning by the use of superspace  \cite{6}-\cite{9}.

 BV-formalism has been used for quantizing  $W_3$ gravity \cite{10}-\cite{12}. 
It has also been used in quantizing metric-affine gravity in two dimentions \cite{13}. 
 The BRST symmetry  for perturbative quantum gravity in four-dimensional flat spacetime
 have been studied by a number of  authors \cite{14}-\cite{16} and their 
work has been summarized by N. Nakanishi and I. Ojima \cite{17}.
 The BRST symmetry in two-dimensional curved  spacetime has 
 been  studied  thoroughly \cite{18}-\cite{20}. Similarly, the BRST symmetry 
for topological quantum gravity in  curved spacetime  \cite{21,22} and the
BRST symmetry for perturbative quantum gravity in both linear and  non-linear gauge's has 
also been analysed \cite{f}.  However, so far no work has been done in analysing the 
non-anticommutative  perturbative quantum gravity in superspace BV-formalism.
This is what we aim to do in this paper.  

\section{Perturbative Quantum Gravity }
We shall analyse perturbative quantum gravity with the following 
hyperbolic complex metric \cite{299p},
\begin{equation}
g_{bc}^{(f)}=b^{(f)}_{bc}+ \omega a^{(f)}_{bc},
\end{equation}
where $\omega$ is the pure imaginary element of a hyperbolic complex
Clifford algebra  with   $\omega^2=+1$. Here 
 $\omega$ forms a ring of numbers and not
a field which the usual system of complex numbers do.
 The advantage of using $\omega$ is that the 
negative energy states
coming from the purely imaginary part of the metric will be avoided. 

Now we can introduce  non-anticommutativity   as follows, 
\begin{eqnarray}
[\hat{x}^a ,\hat{x}^b]&=&2\hat{x}^a \hat{x}^b +i\omega\tau^{ab},
\end{eqnarray}
where $\tau^{ab}$ is a  symmetric tensor. 
We use Weyl
ordering and  express the Fourier transformation of this metric  as, 
\begin{equation}
\hat{g}^{(f)}_{ab} (\hat{x}) =
\int d^4 k \pi e^{-i k \hat{x}  } \;
g^{(f)}_{ab}  (k).
\end{equation}
Now  we  have a one to one map between a function of
$\hat{x}$ to a function of ordinary
coordinates $ y$ via
\begin{equation}
g^{(f)}_{ab} (x)  =
\int d^4 k \pi e^{-i k x } \;
g^{(f)}_{ab} (k).
\end{equation}
So, the  product of  ordinary functions  is given by 
\begin{equation}
g^{(f)ab}(x)\Diamond g^{(f)}_{ab}(x) =   \exp \frac{\omega}{2} (
\tau^{ab} \partial _a^2 \partial _b^1 )
 g^{(f)ab} (x_1) g^{(f)}_{ab}  (x_2)
\left. \right|_{x_1=x_2=x}.
\end{equation}
 Now  $R^{(f)a}_{bcd}$ given as,
\begin{equation}
R^{(f)a}_{bcd}=-\partial _d \Gamma^a_{bc} +\partial _c\Gamma^a_{bd}+\Gamma^a_{e c}\Diamond\Gamma^e_{bd}
-\Gamma^a_{ed}\Diamond\Gamma^e_{bc},
\end{equation}
and we also get  $R_{bc}=R^d_{bcd}$.
Thus finally  $R^{(f)}$ is given by  
\begin{equation}
 R^{(f)} = g^{(f)ab}\Diamond R_{ab}^{(f)}.
\end{equation}
 The Lagrangian density for pure gravity with  cosmological constant $\lambda$ can now be written as, 
\begin{equation}
 \mathcal{L}_c  = \sqrt{g}^{(f)}\Diamond(R^{(f)} - 2\lambda),
\end{equation}
where we have adopted  units, such that  $16 \pi G = 1 $.
In perturbative gravity on flat spacetime one splits the full metric $g_{ab}^{(f)}$ into $\eta_{ab}$ which is the metric for the
 background flat spacetime  and  $h_{ab}$ which is a small perturbation around the
 background spacetime,
\begin{equation}
 g_{ab}^{(f)} = \eta_{ab} + h_{ab}.
\end{equation}
Here both $\eta_{ab}$ and $h_{ab}$ are complex. 
The covariant derivatives along with the lowering and raising of indices are compatible with the 
metric for the background spacetime. The small perturbation $h_{ab}$ is viewed as the field that is to be quantized.
\section{BV-Formalism}
All the degrees of freedom in $h_{ab}$ are not physical as the  Lagrangian density for $h_{ab}$
 is invariant under the following 
 gauge transformations, 
\begin{eqnarray}
\delta_\Lambda h_{ab} &=&  D^e_{ab}\Diamond \Lambda_e \nonumber \\ &=& [ \delta^e_b \partial _a  + \delta^e_a \partial  _b  
+ g^{ce} \Diamond (\partial _c h_{ab}) + \nonumber \\ &&g^{ec} \Diamond h_{ac}\partial  _b  +\eta^{ec}\Diamond h_{cb} \partial  _a ]
\Diamond \Lambda_e. \label{eq}
\end{eqnarray}
In order to remove these unphysical  degrees of freedom, we need to fix a gauge 
by adding a gauge-fixing term along with a ghost term. 
In  the most general covariant gauge 
 the sum of the gauge-fixing term and the ghost term can be expressed as, 
\begin{equation}
\mathcal{L}_g  =  s \Psi,
\end{equation}
where 
\begin{equation}
\Psi =   \overline{c}^a \Diamond \left(\partial ^b h_{ab} - k \partial _a h + \frac{1}{2} b_a \right), \label{c} 
\end{equation}
with $k \neq 1$. Now the sum of the ghost term, the gauge-fixing term 
and the  original classical Lagrangian density 
is invariant under the following BRST
transformations
\begin{eqnarray}
s \,h_{ab} = D^e_{ab}\Diamond c_e,  &&
s \,c^a - c_b \Diamond \partial ^b c^a, \nonumber \\
s \,\overline{c}^a = - b^a, &&
s \,b^a =0.
\end{eqnarray}

BV-formalism  is used to analyse the extended BRST symmetry. This 
 extended BRST symmetry for perturbative quantum gravity can be obtained by 
 first shifting  all the original fields as, 
\begin{eqnarray}
 h_{ab} \to h_{ab} -\tilde{h}_{ab}, &&
 c^a \to c^a -\tilde{c}^a, \nonumber \\ 
\overline{c}^a \to \overline{c}^a -\tilde{ \overline{c}}^a,&&
b^a \to b^a -\tilde{b}^a,
\end{eqnarray}
and then requiring  
the resultant  theory to be invariant under both 
the original BRST transformations and these shift transformations. 
This can be achieved by letting the original fields transform as, 
\begin{eqnarray}
s \,h_{ab} =\psi_{ab}, &&
s \,c^a = \phi^a , \nonumber \\
s \,\overline{c}^a  =\overline{\phi}^a, &&
s \,b^a =\rho^a,
\end{eqnarray}
and the shifted fields  transform as 
\begin{eqnarray}
s \,\tilde{h}_{ab} =\psi_{ab}- {D'}  ^e_{ab} \Diamond (c_e - \tilde{c}_e),&&
s \,\tilde{c}^a = \phi^a+ ( c_b  -\tilde{c}_b) \Diamond \partial ^b (c_a- \tilde{c}_a),  \nonumber \\
s \,\tilde{\overline{c}}^a =\overline{\phi}^a + (b^a-\tilde{b}^a),  && 
s \,\tilde{b}^a =\rho^a.
\end{eqnarray}
Here $\psi_{ab}, \phi^a, \overline{\phi}^a, $ and $  \rho_a$
 are ghosts associated with the shift symmetry and their BRST transformations vanish,
\begin{equation}
  s\, \psi_{ab} = s\, \phi^a = \, s\,\overline{\phi}^a = s\,  \rho^a = 0.
\end{equation}
We  define antifields with opposite parity corresponding to all the original fields.
 These antifields have the following BRST transformations, 
\begin{eqnarray}
s \,h^*_{ab} =\overline{b}_{ab}, &&
s \,c^{*a} = B^a, \nonumber \\
s \,\overline{c}^{*a}  =\overline{B}^a, &&
s \,b^{*a} =\overline{b}^{a}.
\end{eqnarray}
Here $\overline{b}_{ab},  B^a, \overline{B}^a, $ and $ \overline{b}^{a}$  
are Nakanishi-Lautrup fields and their BRST transformations  vanish too,
\begin{equation}
s\, \overline{b}_{ab} = s\, B^a = s\, \overline{B}^a = s\, \overline{b}^a =0.
\end{equation}
 It is useful to define 
\begin{eqnarray}
h'_{ab} = h_{ab}- \tilde{h}_{ab},  &&
c'^a = c^a - \tilde{c}^a , \nonumber \\
\overline{c}'^a  =\overline{c}^a - \overline{\tilde{c}}^a, &&
b'^a =b^a - \tilde{b}^a.
\end{eqnarray}
The physical requirement for the sum of the gauge-fixing term and 
the ghost term is that
all the fields associated with shift symmetry vanish. 
This can be achieved by choosing the following Lagrangian density,  
\begin{eqnarray}
\tilde{\mathcal{L}}_g &=& -\overline{b}_{ab}  \Diamond \tilde{h}^{ab} - h^{*ab} \Diamond (\psi_{ab} - {D'}  ^e_{ab} \Diamond (c_e')) \nonumber \\
&&- \overline{B}^a\Diamond \tilde{c}_a + \overline{c}^*_a \Diamond \left( \phi^a +  ( c'_b) \Diamond  \partial ^b ({c^a}') \right)\nonumber \\
&&+ B^a\Diamond \tilde{\overline{c}}_a - c^*_a \Diamond \left(\overline{\phi}^a +  ({b^a}')\right) + B^a\Diamond
\tilde{b}_a +b^{*a}\Diamond \rho_a. \label{a} 
\end{eqnarray}
The integrating out  the Nakanishi-Lautrup fields in this 
Lagrangian density will make all the shifted fields   vanish. 

If we choose a gauge-fixing fermion
 $\Psi$, 
such that it depends only on the original fields  and furthermore define 
 $\mathcal{L}_g  = s \Psi$, then  we have 
\begin{eqnarray}
 \mathcal{L}_g  
&=&  -\frac{\delta \Psi}{\delta h_{ab}}\Diamond \psi_{ab} + \frac{\delta \Psi}{\delta c_{a}}\Diamond
\phi_a +  \frac{\delta \Psi}{\delta \overline{c}_{a}}\Diamond \overline{\psi}_a - 
 \frac{\delta \Psi}{\delta b_{a}} \Diamond \rho_a. \label{b}
\end{eqnarray}
The total Lagrangian density is given by 
\begin{equation}
  \mathcal{L} = \mathcal{L}_c (h- \tilde{h} ) + \tilde{\mathcal{L}}_{g} + \mathcal{L}_g. 
\end{equation}
After integrating out the  Nakanishi-Lautrup fields, 
this total Lagrangian density can be written as, 
\begin{eqnarray}
 \mathcal{L} &=& \mathcal{L}_c (h- \tilde{h} ) +  h_{ab}^* \Diamond D_e^{ab}c^e + \overline{c}^{*a} \Diamond c^b \Diamond \partial _b c_a \nonumber \\
&& - c^{*a} \Diamond b_a - \left(h_{ab}^* +\frac{\delta \Psi}{\delta h^{ab}} \right)\Diamond \psi^{ab} + 
  \left(\overline{c}_{a}^* +\frac{\delta \Psi}{\delta c ^{a}} \right) \Diamond \phi^a \nonumber \\
&&  - \left(c_{a}^* -\frac{\delta \Psi}{\delta\overline{c}^{a}} \right)\Diamond \overline{\phi}^a  + 
 \left(b_{a}^* -\frac{\delta \Psi}{\delta b^{a}} \right)\Diamond \rho^a.
\end{eqnarray}
Now integrating 
out the ghosts associated with the shift symmetry, 
we get the following expression for the antifields, 
\begin{eqnarray}
h^*_{ab} =-\frac{\delta \Psi}{\delta h^{ab}},&&
c^{*a} = \frac{\delta \Psi}{\delta \overline{c}^{a}} , \nonumber \\
\overline{c}^{*a}  =-\frac{\delta \Psi}{\delta c^{a}}, &&
b^{*a} =\frac{\delta \Psi}{\delta b^{a}}.
\end{eqnarray}
These equations along with Eq. ($\ref{c}$) fix the exact expressions for the antifields in terms of the original fields. 
\section{Superspace Formulation} 
In this section we will express the results of the previous section in superspace formalism with one anti-commutating  variable. 
Let $\theta$ be an anti-commutating variable, then we can define the following superfields, 
\begin{eqnarray}
 \omega_{ab} = h_{ab} + \theta \psi_{ab}, &&
 \tilde{\omega}_{ab} = \tilde{h}_{ab} + \theta (\psi_{ab} - {D'}  _{ab}^e \Diamond (c_e')),\nonumber \\
\eta_a = c_a +\theta \phi_a, &&
\tilde{\eta}_a = \tilde{c}_a + \theta(\phi_a +( c_b') \Diamond \partial ^b (c_a') ), \nonumber \\
\overline{\eta}_a =c_a +\theta \overline{\phi}_a, &&
\tilde{\overline{\eta}}_a = \tilde{\overline{c}}_a + \theta(\overline{\phi}_a +( b_a' )), \nonumber \\
f_a = b_a +\theta \rho_a, &&
\tilde{f}_a = \tilde{b}_a +\theta \rho_a,
\end{eqnarray}
and the following anti-superfields, 
\begin{eqnarray}
 \tilde{\omega}^*_{ab} = h^*_{ab} - \theta \overline{b}_{ab},&&
\tilde{\eta}^*_a = c^*_a - \theta B_a, \nonumber \\
\tilde{\overline{\eta}}^*_a=\overline{c}^*_a - \theta \overline{B}_a, &&
\tilde{f}^*_a = b_a^*  - \theta \overline{b}_a. 
\end{eqnarray}
From these two equations, we have 
\begin{eqnarray}
 \frac{\partial }{ \partial \theta}    \tilde{\omega}^{*ab}\Diamond\tilde{\omega}_{ab} &=& - \overline{b}_{ab}\Diamond \tilde{h}^{ab} -h^{*ab} \Diamond
(\psi_{ab} - {D'}  ^e_{ab}\Diamond
(c_e')), \nonumber \\
 \frac{\partial }{ \partial \theta}   \tilde{\overline{\eta}}^*_a \Diamond \tilde{\eta}^a  &=&  -\overline{B}^a\Diamond
 \tilde{c}_a + \overline{c}^{*a} \Diamond (\psi_a  
  + ( c_b') \Diamond \partial ^b (c_a')),\nonumber \\
 -\frac{\partial }{ \partial \theta}   \tilde{\overline{\eta}}^a \Diamond \tilde{\eta}^*_a  &=& B^a \Diamond \tilde{\overline{c}}_a  - 
c^*_a \Diamond ( \overline{\phi}^a  + ({b^a}')), \nonumber \\
- \frac{\partial }{ \partial \theta}    \tilde{f}^*_a \Diamond\tilde{f}^a  &=& \overline{b}^a \Diamond \tilde{b}_a + b^{*a}\Diamond  \rho_a.
\end{eqnarray}
Now we can express  $\tilde{\mathcal{L}}_g $  given by Eq. ($\ref{a}$) as,
\begin{eqnarray}
 \tilde{\mathcal{L}}_g &=& \frac{\partial }{ \partial \theta}   (\tilde{\omega}^{*ab}\Diamond
\tilde{\omega}_{ab} +\tilde{\overline{\eta}}^*_a \Diamond \tilde{\eta}^a - 
\tilde{\overline{\eta}}^a \Diamond \tilde{\eta}^*_a - \tilde{f}^*_a \Diamond\tilde{f}^a ).
\end{eqnarray}
Furthermore, if  we define $\Psi$ as, 
\begin{eqnarray}
 \Phi &=& \Psi + \theta s \Psi\nonumber \nonumber \\&=& \Psi + \theta \left (-\frac{\delta \Psi}{\delta h_{ab}}\Diamond
\psi_{ab} + \frac{\delta \Psi}{\delta c_{a}}\Diamond\phi_a +
  \frac{\delta \Psi}{\delta \overline{c}_{a}}\Diamond\overline{\psi}_a - 
 \frac{\delta \Psi}{\delta b_{a}} \Diamond \rho_a\right), 
\end{eqnarray}
then we can express $\mathcal{L}_g$  given by Eq. ($\ref{b}$) as,
\begin{equation}
 \mathcal{L}_g =  \frac{\partial }{ \partial \theta} \Phi  .
\end{equation}
Now the complete Lagrangian density in the superspace formalism is given by, 
 \begin{eqnarray}
  \mathcal{L} &=&  \frac{\partial }{ \partial \theta} \Phi    + \frac{\partial }{ \partial \theta}   
(\tilde{\omega}^*_{ab}\Diamond \tilde{\omega}^{ab} + 
\tilde{\overline{\eta}}^*_a \Diamond \tilde{\eta}^a-\tilde{\overline{\eta}}^a\Diamond \tilde{\eta}_a^* -
\tilde{\overline{\eta}}^a\Diamond \tilde{\eta}_a^* - \tilde{f}^*_a \Diamond \tilde{f}^a) \nonumber \\ && 
+\mathcal{L}_c( h_{ab} -\tilde{h}_{ab}).
 \end{eqnarray}
Upon elimination of the Nakanishi-Lautrup fields and 
the ghosts associated with shift symmetry,  this Lagrangian density
 is manifestly invariant under the BRST symmetry as well as the shift symmetry. 
\section{Conclusion}
In this paper we  analysed non-anticommutative perturbative gravity with a complex metric.
As this theory contained unphysical  degrees of freedom,  we  added
 a gauge-fixing term and a ghost term to it. 
We found that the sum of the original classical Lagrangian density, 
the gauge-fixing term and the ghost term was invariant under the BRST transformations. 
As the shifting of fields occurs naturally in the background field method, 
we analysed the effect of the shift symmetry in the BV-formalism. 
Finally, we expressed our results in the superspace formalism using a single Grassmann parameter.  

 It is well known that the sum of the original classical Lagrangian density,
 the gauge-fixing term and the ghost term for 
most  theories posing BRST symmetry 
is also invariant under another symmetry called the anti-BRST symmetry \cite{17}. 
It will be interesting 
to investigate the anti-BRST version of this theory. Furthermore,
 the invariance of a gauge theory under  the
BRST and the anti-BRST transformations along with the shift transformations has already been 
 analysed in the superspace BV-formalism \cite{6}. 
Thus, after analysing the anti-BRST symmetry  for this theory, the invariance of this theory under the    
original BRST and the original anti-BRST transformations along with shift transformations can be studied in the 
superspace BV-formalism. 

It will also be interesting to generalise the results of this paper to general curved spacetime.
The generalisation to arbitrary spacetime might not be so simple as it is still not  completely clear
 how the BRST symmetry  works for general curved spacetime. There will also be ambiguities due to 
the definition of  vacuum state. We know  it is possible to define a vacuum state called 
the Euclidean vacuum in maximally symmetric spacetime \cite{at}. We  also know  
  the ghosts in   anti-de Sitter  spacetime do not 
contain any infrared divergence. Therefore, the generalisation of this work  to anti-de Sitter  
spacetime can be done easily  \cite{fa}. However, as
 the ghosts in  de Sitter spacetime contain infrared divergence,
 this work can not be directly extended to de Sitter spacetime 
\cite{fa}. In order to   generalize this work to de Sitter spacetime we will have to modify
 the BRST transformations accordingly. 

\end{document}